\documentclass[review]{elsarticle}
\usepackage[utf8]{inputenc}
\usepackage[T1]{fontenc}
\usepackage{amsmath,amsfonts,amssymb,amsthm,bm,color,dsfont,graphicx,psfrag,mathtools,hyperref,float,subcaption}

\begin{document}
\title{Route to chaos in a branching model of neural network dynamics}

\author{Rashid V. Williams-García\corref{cor1}}
\ead{Rashid.Williams-Garcia@lmpt.univ-tours.fr}
\author{Stam Nicolis}
\affiliation{organization={Institut Denis Poisson, Université de Tours, Université d'Orléans, CNRS (UMR7013)},
addressline={Parc de Grandmont},
postcode={F-37200},
city={Tours},
country={France}}

\cortext[cor1]{Corresponding author}

\date{\today}

\begin{abstract}
\noindent Simplified models are a necessary steppingstone for understanding collective neural network dynamics, in particular the transitions between different kinds of behavior, whose universality can be captured by such models, without prejudice. One such model, the cortical branching model (CBM), has previously been used to characterize part of the universal behavior of neural network dynamics and also led to the discovery of  a second, chaotic transition which has not yet been fully characterized. Here, we study the properties of this chaotic transition, that occurs in the mean-field approximation to the $k_{\sf in}=1$ CBM  by focusing on the constraints the model imposes on initial conditions, parameters, and the imprint thereof on the Lyapunov spectrum. Although the model seems similar to the Hénon map, we find that the Hénon map cannot be recovered using orthogonal transformations to decouple the dynamics. Fundamental differences between the two, namely that the CBM is defined on a compact space and features a non-constant Jacobian, indicate that the CBM maps, more generally,  represent a class of generalized Hénon maps which  has yet to be fully understood.
\end{abstract}


\maketitle

\section{Introduction}\label{intro}

\noindent Years of experimental evidence have led to the hypothesis that healthy brain networks tune themselves to operate in a ``quasicritical'' region~\cite{mora2011biological,williams2014quasicritical}. And while alternative explanations for the data have been proposed \cite{touboul2017power}, the idea of a critical point guiding brain function remains intriguing, as in its vicinity, brain function  would optimize information processing, storage capacity, and task flexibility \cite{shew2011information}. Moreover, universal properties of the transition would eclipse the relevance of the  details of the properties  of individual neurons and their interactions \cite{stanley1987introduction}, meaning that  certain features of brain dynamics could be described by much simpler models. The cortical branching model (CBM) is one such model which has successfully reproduced many dynamical features of living brain tissue \cite{haldeman2005critical, pajevic2009efficient, chen2010few}. In addition, the CBM has been instrumental in the development of the non-equilibrium Widom line framework which describes how the quasicritical region changes based on external inputs \cite{williams2014quasicritical}. As these changes unfold, neural networks have to adjust their properties (\textit{e.g.}, the strength of their synaptic connections) to remain in the quasicritical region, potentially navigating a complex phase diagram in the process. This idea is already serving as an organizing principle to explain existing experimental results \cite{fosque2021evidence}.

In the mean-field approximation of the CBM, three phases and two phase transitions have already been observed: phases with and without sustained activity in the absence of external stimulation (\textit{i.e.}, the ordered and disordered phases, respectively) separated by a second-order  phase transition, and an unexplored transition to a chaotic phase, previously termed ``quasiperiodic'' in \cite{williams2014quasicritical}. The approximation leads to the study  of a multi-dimensional non-linear discrete map; certain constraints imposed on this map induce folding and mixing, thus resulting in chaotic behavior for certain parameter regions. Many-body simulations of the CBM within the chaotic phase have revealed synchronous bursting patterns, a phenomenon which has long been associated with epileptic seizures \cite{jiruska2013synchronization}. The exact relationship between chaos and brain function, however, remains mysterious. An interesting example involves electroencephalogram (EEG) recordings, which have revealed chaotic activity in both healthy individuals and, to a lesser extent, individuals diagnosed with schizophrenia and upon seizure onset in epileptics \cite{kim2000estimation, sackellares2000epilepsy}. In such systems, the degree of sensitivity to initial conditions (quantified by the Lyapunov exponents) is often used to deduce the presence of chaos \cite{nicolis1985chaotic,hasselblatt2003first}. The fact that these EEG recordings show positive Lyapunov exponents in healthy cases and lower (but still positive) Lyapunov exponents in disease cases potentially brings into focus the role of the chaotic transition and its relationship to the critical transition, implying an intricate relationship between chaos, quasicriticality, and brain function. A more detailed examination of the CBM's chaotic phase could help understand this relationship.

To gain intuition into how chaos emerges, we use the quadratic mean-field approximation of the CBM studied in \cite{williams2014quasicritical} to examine the route to chaos and the structure of the chaotic phase for the special case of a network whose nodes can be in  two states. For this case, analytical calculations become feasible, in particular pertaining to the determination of fixed points and their stability. Of particular interest is that the properties of the dynamical variables (which take values between 0 and 1) imply constraints on the initial conditions as well as on the parameters of the map, analogous to those relating the Julia and Mandelbrot sets. Furthermore, while the map may bear some resemblance to the Hénon map \cite{mosekilde2000bifurcation}, we provide evidence based on the calculation of the spectrum of Lyapunov exponents that the constraints define a distinct universality class.

\section{The Cortical Branching Model in the Mean--Field Approximation}\label{CBMMFA}

\noindent In the CBM, neurons are represented by nodes interacting on a directed network while neuronal membrane potentials are represented by dynamical node states $z\in\{0,1,2,...,\tau_{\sf r}\}$. The state $z=0$ corresponds to a resting potential (quiescence), $z=1$ to a depolarizing action potential (activation), and higher values to a post-activation refractory period, where the higher the value $\tau_{\sf r}$, the more extensive the hyperpolarization and thus the longer the refractory period. Quiescent nodes become active by means of external stimuli (represented in the CBM as spontaneous activation with probability $p_{\sf s}$) or can be driven to activation by active neighbors with synaptic weights corresponding to probabilities proportional to the branching parameter $\kappa$. Interpreted as the ratio of the number of descendant activations to the number of ancestor activations, the branching parameter $\kappa$ quantifies the spread of activity in the network. If $\kappa<1$, network activity tends to recede, whereas if $\kappa>1$, network activity tends to spread. Following activation, a node's state cycles through  states $z=2$ up to $z=\tau_{\sf r}$, after which the node returns to quiescence.

Phases of the CBM are characterized in the mean-field approximation by an order parameter, $\vec{x}\equiv(x_1,\ldots,x_{\tau_{\sf r}}),$ whose $\tau_{\sf r}$ components, $x_i^*$, represent the average collective activation rate, \textit{i.e.}, the mean firing rate of the neurons.  The three phases include a disordered phase where $\vec{x}^*={\bf 0}$, an ordered phase with $0<x_i^*\leq1$, and a so-called \textit{quasiperiodic} phase where the fixed point $\vec{x}_i^*$ is unstable. Within the \textit{quasiperiodic} phase, the density of active nodes $x_1$ oscillates continuously, never reaching the fixed point $x_1^*$: a behavior which has been reproduced in simulations \cite{williams2014quasicritical}. Although its significance for brain function remains unknown, it was suggested that the \textit{quasiperiodic} phase could represent a pathological state corresponding to a neurological disorder. However it could be argued that chaotic behavior within the \textit{quasiperiodic} phase could, in fact, describe a healthy system, since it would then be much {\em more} robust to external perturbations and better in adapting to them~\cite{goldberger1991normal}. Previously, the distinction between purely periodic and chaotic behavior in the CBM was not studied in depth, and so to better understand its role in the function of neural networks, we carry out a further examination of the \textit{quasiperiodic} phase and transitions to it in the mean-field approximation.

The mean-field approximation of the CBM is given by a $\tau_{\sf r}$-dimensional discrete map $\vec{x}_{n+1}=F(\vec{x}_n)$, where $\tau_{\sf r}$ represents the integer-valued neuronal refractory period and $F$ is a vector-valued function \cite{williams2014quasicritical}. As a first example, we consider the case that each node has an in-degree of 1 and $\tau_{\sf r}=2$, wherein $F$ has components
\begin{equation}
	\label{CBMtaur=2}
	\begin{array}{l}
		\displaystyle
		x_{n+1}=F_1(\vec{x}_n)\equiv(1-x_n-y_n)\left(\kappa(1-p_{\sf s})x_n+p_{\sf s}\right)\\
		\displaystyle
		y_{n+1}=F_2(\vec{x}_n)\equiv x_n,
	\end{array}
\end{equation}
where $0\leq x_n\leq1$ and $0\leq y_n\leq1$ represent the density of nodes in states 1 and 2, respectively, at time-step $n$. We use the $(x_n, y_n)$ notation here for readability, although we note that the notation $x_{z, n}$ is better suited for larger values of $\tau_{\sf r}$. Set in these terms, these equations define a two-dimensional dynamical system with quadratic non-linearity; it is therefore expected to show a transition from regular to chaotic behavior, similar to that of the Hénon map; it is this onset which we are interested in. 

From experience with the Julia and Mandelbrot sets \cite{lauwerier1991fractals}, we expect that there are two points of interest: (a) the admissible initial conditions $(x_0,y_0)$ for which the orbit remains in the unit square $[0,1]\times[0,1]$; and (b) the values of the parameters $\kappa$ and $p_{\sf s}$ which ensure that admissible initial conditions result in orbits which stay within the unit square. The difference in the present case is  that the CBM map is non-holomorphic---a feature which complicates the analysis considerably. In the following we shall, therefore, study the constraints on the initial conditions and subsequently show what these imply for the admissible region in the $\kappa$--$p_{\sf s}$ plane. 

\subsection{Admissible initial conditions}\label{valinitcond}
 \noindent A major challenge when calculating orbits of this map involves determining the initial conditions $\vec{x}_0$ from which the subsequent evolution of the map remains within the square $[0,1]\times[0,1]$; a problem which bears resemblance to the Julia set \cite{lauwerier1991fractals}. Any of these \textit{admissible} initial conditions must first satisfy the following condition to ensure that $0\leq x_1\leq1$ and $0\leq y_1\leq1$:
$$
	1-\frac{1-x_0}{cx_0+p_{\sf s}}\leq x_0+y_0\leq 1+\frac{x_0}{cx_0+p_{\sf s}},
$$
where $c\equiv\kappa(1-p_{\sf s})$. We immediately notice that the upper bound is ineffective since it is larger than 1, and so this is reduced to a first criterion for admissible initial conditions:
\begin{equation}
	\label{condition1}
	(1-x_0)\left(1-\frac{1}{cx_0+p_{\sf s}}\right)\leq y_0\leq1-x_0,
\end{equation}
which yields a quadratic lower bound and linear upper bound on $y_0$, outside of which there are no admissible initial conditions. We next determine a natural constraint on the model parameters such that all initial conditions satisfying $0\leq x_0+y_0\leq1$ are admissible by having the lower bound of Eq. \ref{condition1} be less than or equal to 0. This results in $0\leq\kappa\leq1$, a range previously imposed in \cite{williams2014quasicritical}.

\begin{figure}
	\includegraphics[width=\textwidth]{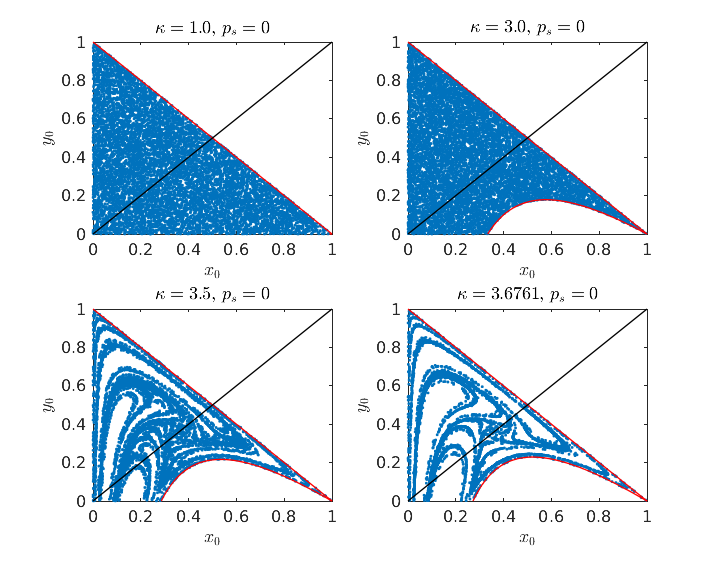}
	\caption{Scatter plots indicating admissible initial conditions for a range of $\kappa$ at $p_{\sf s}=0$ and after $10^6$ iterations. The upper and lower bounds of Eq. \ref{condition1} are shown by the red lines, while the black line indicates the line of potential fixed points of the map. No admissible initial conditions were found for $\kappa>\kappa_\mathrm{max}\approx3.6761$.}
	\label{scatter_ICs0}
\end{figure}

In order to verify our predictions for the admissible initial conditions, we employ a Monte Carlo method, randomly sampling the initial vectors $\vec{x}_0$ within the unit square and iterating Eq. \ref{CBMtaur=2} until the trajectory leaves the unit square. Initial conditions resulting in orbits which remain within the unit square after $10^6$ iterations are recorded, while those which possess any excursions are excluded (such as orbits which escape to infinity). The results have been entered into scatter plots (see Fig. \ref{scatter_ICs0}) and indeed verify the Eq. \ref{condition1} criterion. Importantly, Eq. \ref{condition1} does not capture all inadmissible initial conditions and a fractal structure of admissible initial conditions develops as $\kappa$ increases.

\subsection{Admissible parameter values}\label{valkappaps}
\begin{figure}
	\includegraphics[width=\textwidth]{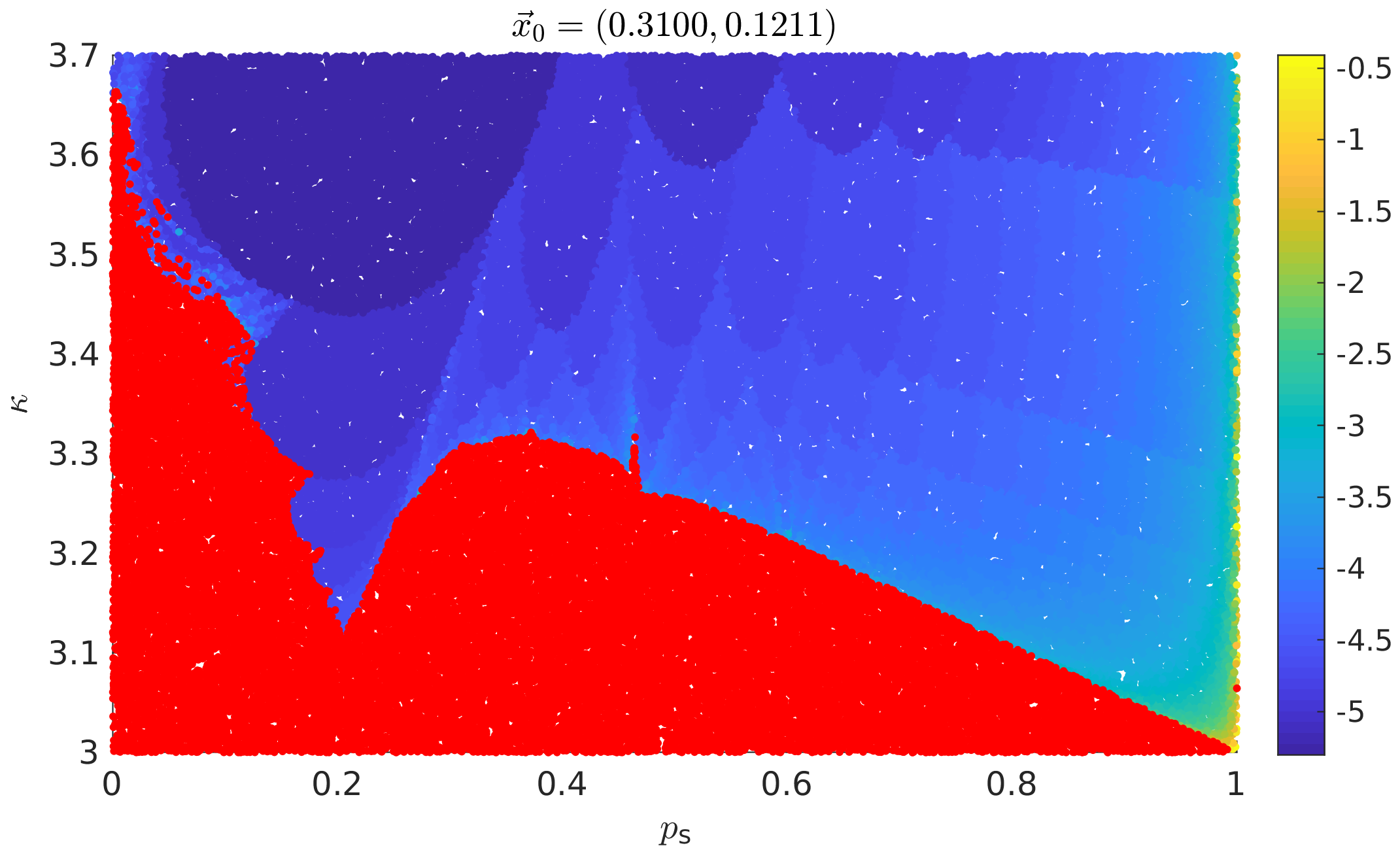}
	\caption{Uniform Monte Carlo sampling on the range of parameter values shown reveals the fractal boundary between admissible (red dots) and inadmissible parameter values (blue-spectrum dots). The color bar shows the logarithm of the fraction of iterations (out of $10^6$) before trajectories with inadmissible parameter values escaped the unit square. The resulting structures result from the fact that not all trajectories leave the unit square at the same rate.}
	\label{MandelCBM}
\end{figure}

\noindent In addition to using only admissible initial conditions, we must restrict ourselves to parameter values which produce orbits remaining within the unit square. By choosing an $\vec{x}_0$ which remained admissible for the entire range of $\kappa\in[0,3.6761]$ and sampling the range of parameter values, we determine the set of admissible $\kappa$ and $p_{\sf s}$ values (cf. Fig. \ref{MandelCBM}), a generalization of the Mandelbrot set. Each choice of $\vec{x}_0$ results in a different set of admissible parameter values.

\section{Results}\label{results}
\subsection{Fixed points and their (in)stability}\label{FPinst}
\noindent Assuming parameter values which define admissible initial conditions and orbits, the CBM map possesses two fixed points of the form $\vec{x}^*_\pm=(x_\pm^\ast,x_\pm^\ast)$, where
\begin{equation}
	\label{fixedpoints}
	x^*_\pm = \frac{(c-2p_{\sf s}-1)\pm\sqrt{(c-2p_{\sf s}-1)^2+8cp_{\sf s}}}{4c}.
\end{equation}
Stability of these $\vec{x}^*_\pm$ is then determined by the eigenvalues of the Jacobian of the CBM map evaluated at the fixed points. For example, when $p_{\sf s}=0$, $\vec{x}^*_-$ is stable for $0\leq\kappa<1$ and $\vec{x}^*_+$ is stable for $1.0\leq\kappa<3.0$, losing stability at $\kappa_\ast=3.0$ for any $p_{\sf s}\in[0,1]$. Periodic orbits appear when $\kappa\geq\kappa_\ast$ with a period-doubling up to $\kappa=\kappa_\mathrm{chaos}^{(1)}\approx3.6740$. This hints at a period-doubling route to chaos, which we explore in more detail in the following.

\begin{figure}
	\centering
	\begin{subfigure}{0.49\textwidth}
		\centering
		\includegraphics[width=\linewidth]{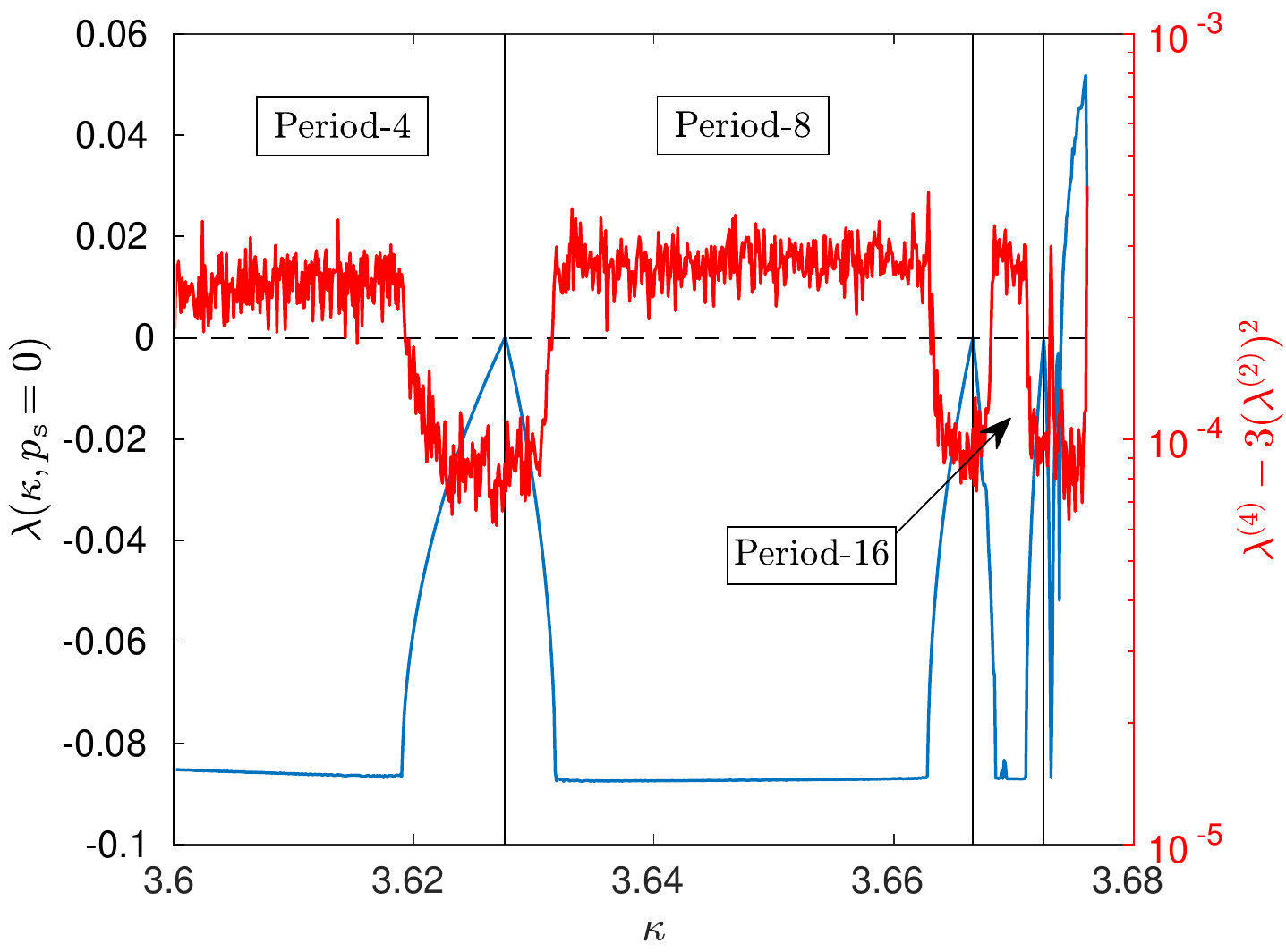}
	\end{subfigure}
	\begin{subfigure}{0.5\textwidth}
		\centering
		\includegraphics[width=\linewidth]{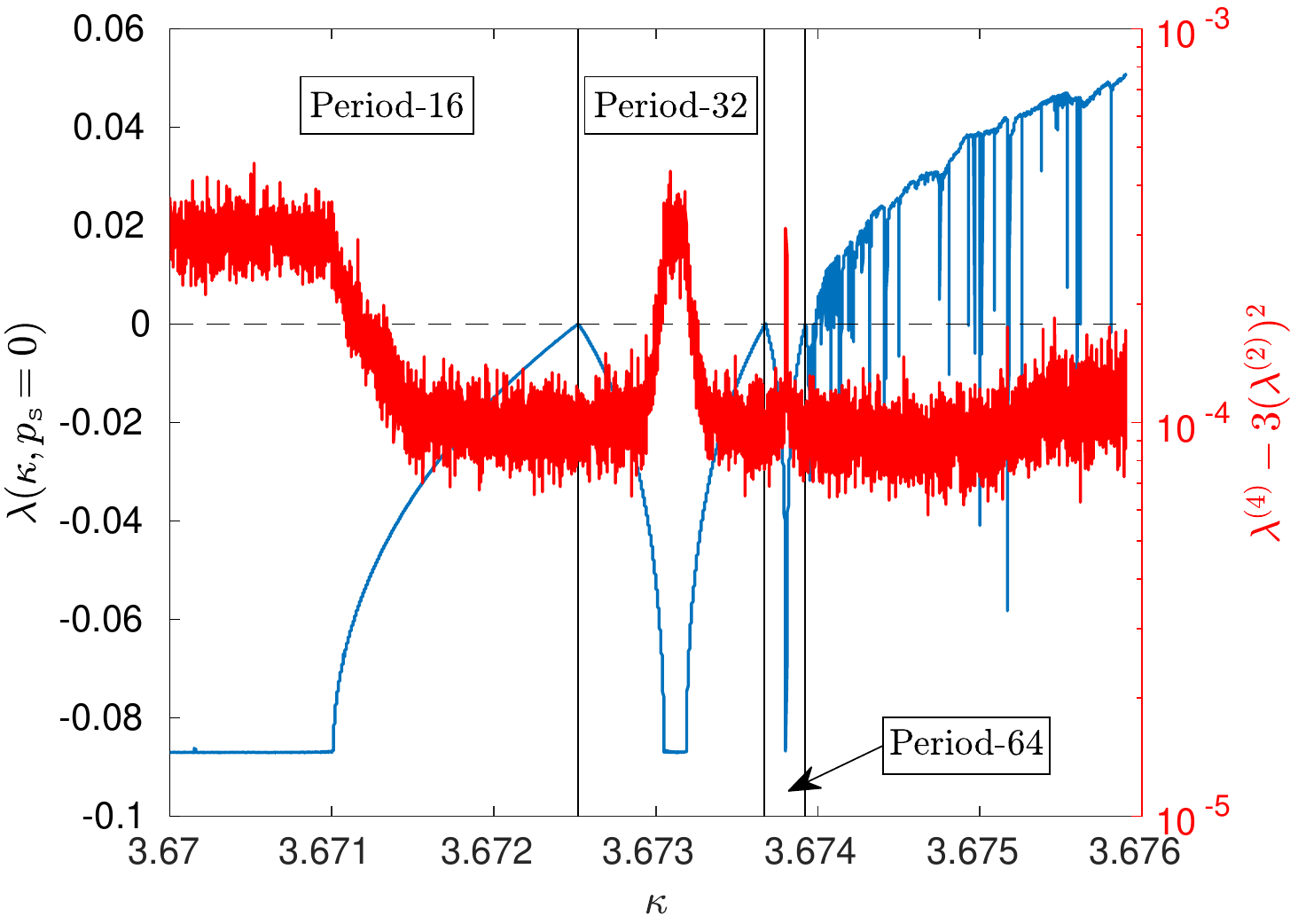}
	\end{subfigure}
	\caption{The $p_{\sf s}=0$ characteristic Lyapunov exponents (blue) averaged over 1000 admissible initial conditions and evaluated numerically after $10^6$ iterations for $\kappa\in[3.6000,3.6761]$ (left) and a zoom into the region $\kappa\in[3.6700,3.6761]$ (right). Lyapunov spectra appear to become more Gaussian near bifurcations (\textit{i.e.}, when $\lambda=0$) and when $\lambda>0$.}
	\label{LyapunovHi}
\end{figure}

\begin{figure}
	\includegraphics[width=\textwidth]{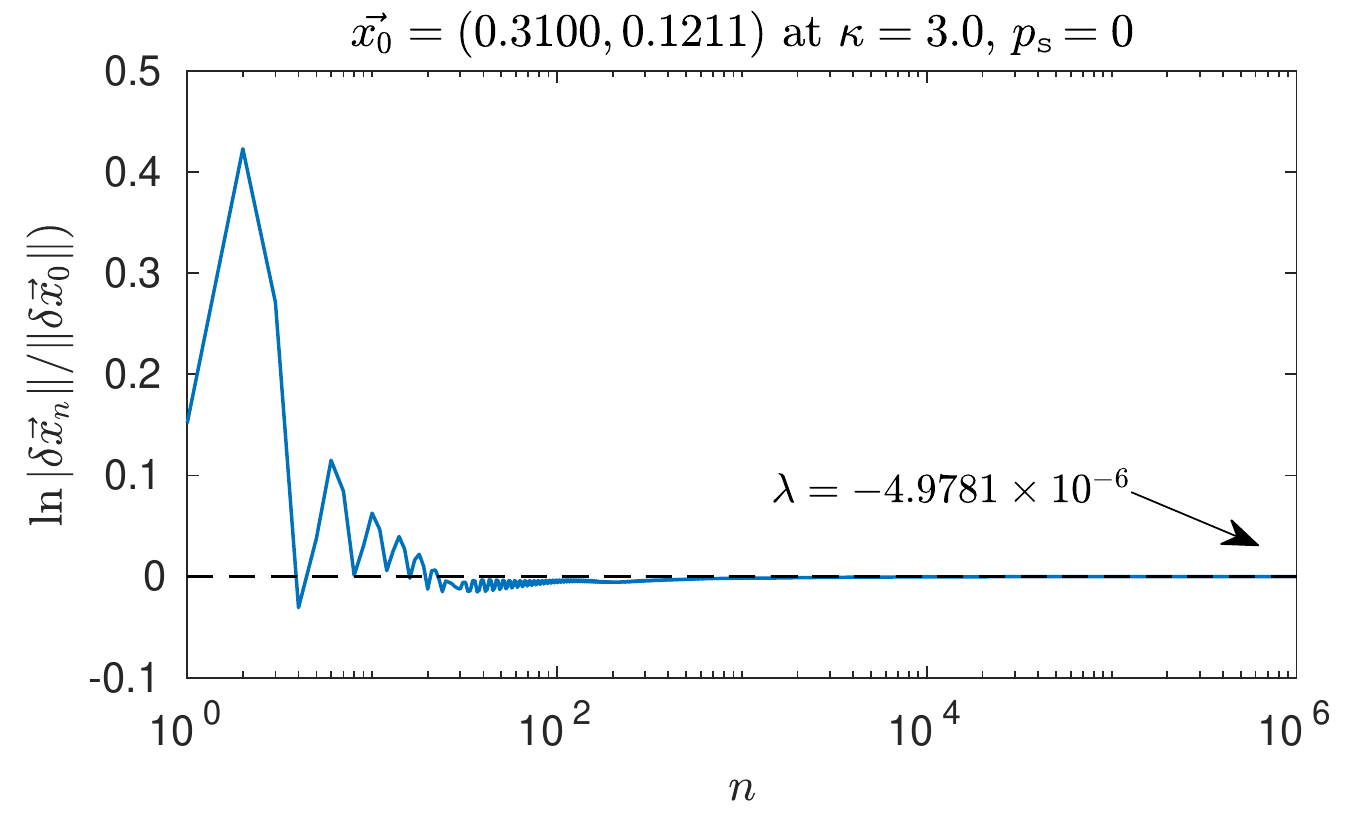}
	\caption{Numerical approximation of the characteristic Lyapunov exponent $lambda$; transient behavior is minimized following $10^4$ iterations.}
	\label{mleSeries}
\end{figure}

Dynamical systems (such as Eq. \ref{CBMtaur=2}) are typically considered chaotic if they feature sensitivity to initial conditions, topological mixing, and dense periodic orbits \cite{hasselblatt2003first}. We utilized the characteristic Lyapunov exponent to quantify sensitivity to initial conditions. If we iterate two copies of the map starting from a pair of initial conditions with deviation $\delta\vec{x}_0$, the characteristic Lyapunov exponent is given by
\begin{equation}\label{Lyapunov1}
    \lambda = \lim_{N\to\infty}\frac{1}{N}\sum_{n=0}^{N-1}\ln\left(\frac{\|\delta\vec{x}_n\|}{\|\delta\vec{x}_0\|}\right),
\end{equation}
where $\delta\vec{x}_n = \prod_{i=0}^{n}DF(\vec{x}_i)\delta\vec{x}_0$ is the deviation after $n$ iterations and $DF(\vec{x}_i)$ is the Jacobian matrix evaluated at $\vec{x}_i$. In most cases, we determine $\lambda$ numerically after removing a transient of $10^4$ iterations, resulting in an approximate value of $\lambda$ which we claim to be representative of the limit because of an observed plateau (cf. Fig. \ref{mleSeries}). However when the map has settled onto a stable fixed point $\vec{x}^*$, $\lambda$ can be determined exactly. In this special case, Eq. \ref{Lyapunov1} reduces to $\lambda^* = \ln|DF(\vec{x}^*)|$, \textit{i.e.}, the natural logarithm of the largest eigenvalue of $DF(\vec{x}^*)$, or
$$
	\lambda^* = \log_2\left(c-p_s-3cx^*+\sqrt{(3cx^*+p_s-c)^2-4(cx^*+p_s)}\right).
$$
For $0<\kappa<1$ and $p_{\sf s}=0$, the fixed point $x^*_-=0$ (cf. Eq. \ref{fixedpoints}) is stable and thus the resulting exponent, $\lambda^*= \ln(\kappa)$ is negative and vanishes at $\kappa=1$. When $1<\kappa<3$ and $p_{\sf s}=0$, the stable fixed point becomes $x^*_+=(\kappa-1)/2\kappa$, giving
$$
	\lambda^* = \log_4\left(3-\kappa+\sqrt{\kappa^2-14\kappa+17}\right),
$$
whose real part vanishes at $\kappa=1$ and $\kappa=3$. We verified this prediction numerically up to $\kappa=3.0$, when the fixed point $x_+^*$ loses stability. We finally demonstrate the onset of chaos by calculating the characteristic Lyapunov exponent up to $\kappa = 3.6761$, observing positive $\lambda$ values at $\kappa=\kappa_\mathrm{chaos}^{(1)}\approx3.6740$ (cf. Fig. \ref{LyapunovHi}). 

\begin{figure}
	\centering
	\includegraphics[width=\textwidth]{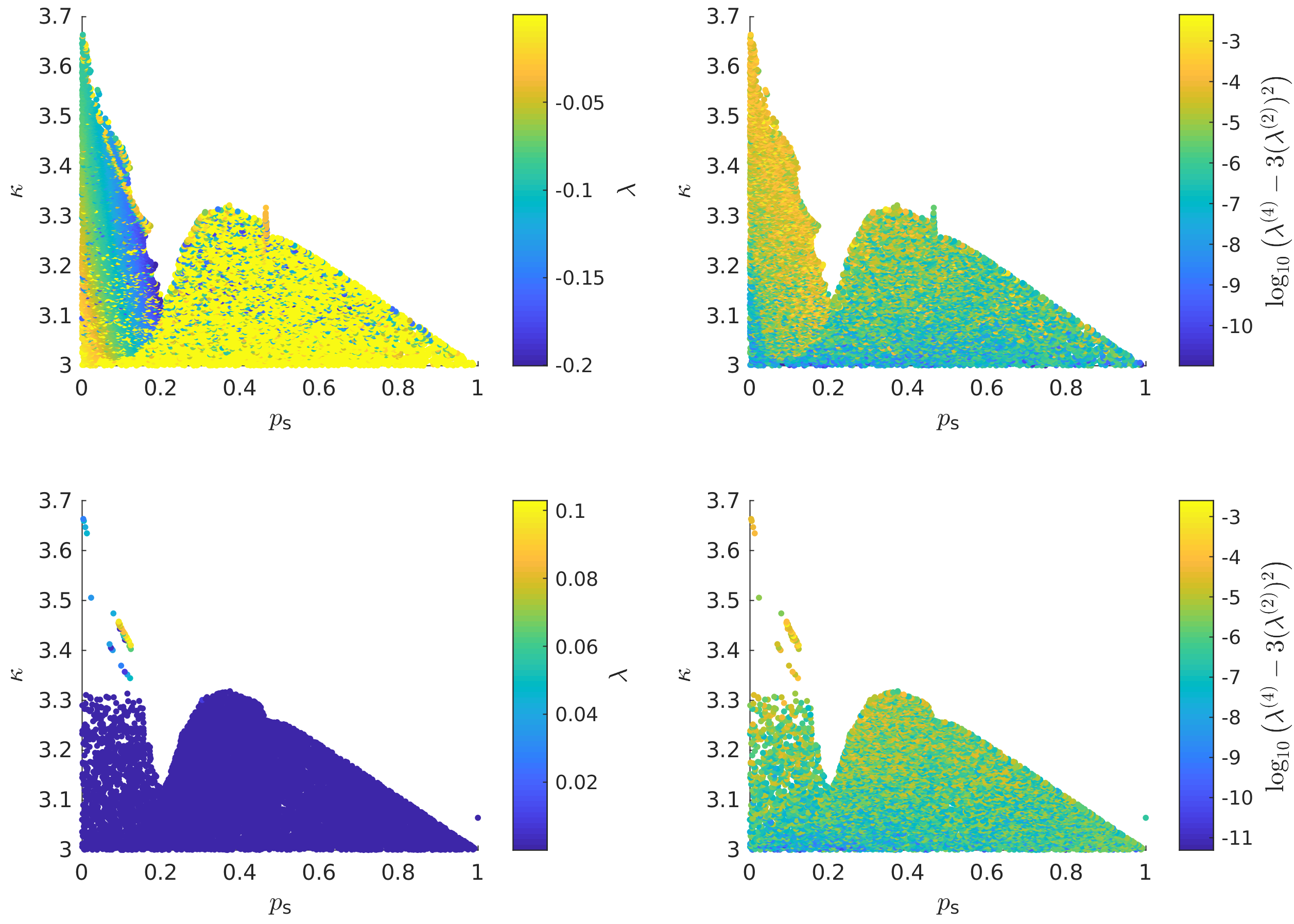}
	\caption{Lyapunov exponents of the CBM illustrate the fractal nature of the chaotic boundaries. While a greater density of chaotic orbits is observed, periodic orbits are still present at high values of $p_{\sf s}$. The same initial condition $\vec{x}_0=(0.3100,0.1211)$ and $(p_{\sf s},\kappa)$ pairs have been used as in Fig. \ref{MandelCBM}.}
	\label{pDiagsJoin}
\end{figure}

Details of the Lyapunov spectrum can be used to further characterize the system. A normal distribution is typically expected, however, non-Gaussian distributions may be associated with special properties of dynamical systems such as intermittency \cite{datta2003non}. We use a cumulant approach to quantify the proximity of the Lyapunov spectrum to a normal distribution following $10^6$ iterations (post-transient) of the CBM. The cumulant is calculated from central moments of the distribution of Lyapunov exponents obtained during those $10^6$ iterations \cite{kenney1951cumulants}. With a Gaussian distribution, we expect that
\begin{equation}
	\label{cumulantEq}
	\lambda^{(4)}-3(\lambda^{(2)})^2 = 0,
\end{equation}
where the $k$th moments are defined as
\begin{equation}
	\nonumber
	\lambda^{(k)} \equiv\langle\lambda^k\rangle = \lim_{N\to\infty}\frac{1}{N}\sum_{n=0}^{N-1}\left(\ln\frac{\|\delta\vec{x}_n\|}{\|\delta\vec{x}_0\|}-\lambda\right)^k.
\end{equation}
We evaluate Eq. \ref{cumulantEq} numerically over a range of $\kappa$ values and observe significant fluctuations of the cumulant expression in the vicinity of bifurcations (cf. Fig. \ref{LyapunovHi}). In order to examine the dependence of the chaotic boundary on $p_{\sf s}$, we have also evaluated the Lyapunov spectra for non-zero $p_{\sf s}$. Our results (cf. Fig. \ref{pDiagsJoin}) illustrate the complexity of the appearance of chaos in the CBM, where pockets of chaotic activity are observed depending on the $\kappa$ and $p_{\sf s}$ values.

\subsection{Along the period--doubling route to chaos}\label{chaos}

\begin{figure}
	\centering
	\begin{subfigure}{0.485\textwidth}
		\centering
		\includegraphics[width=\linewidth]{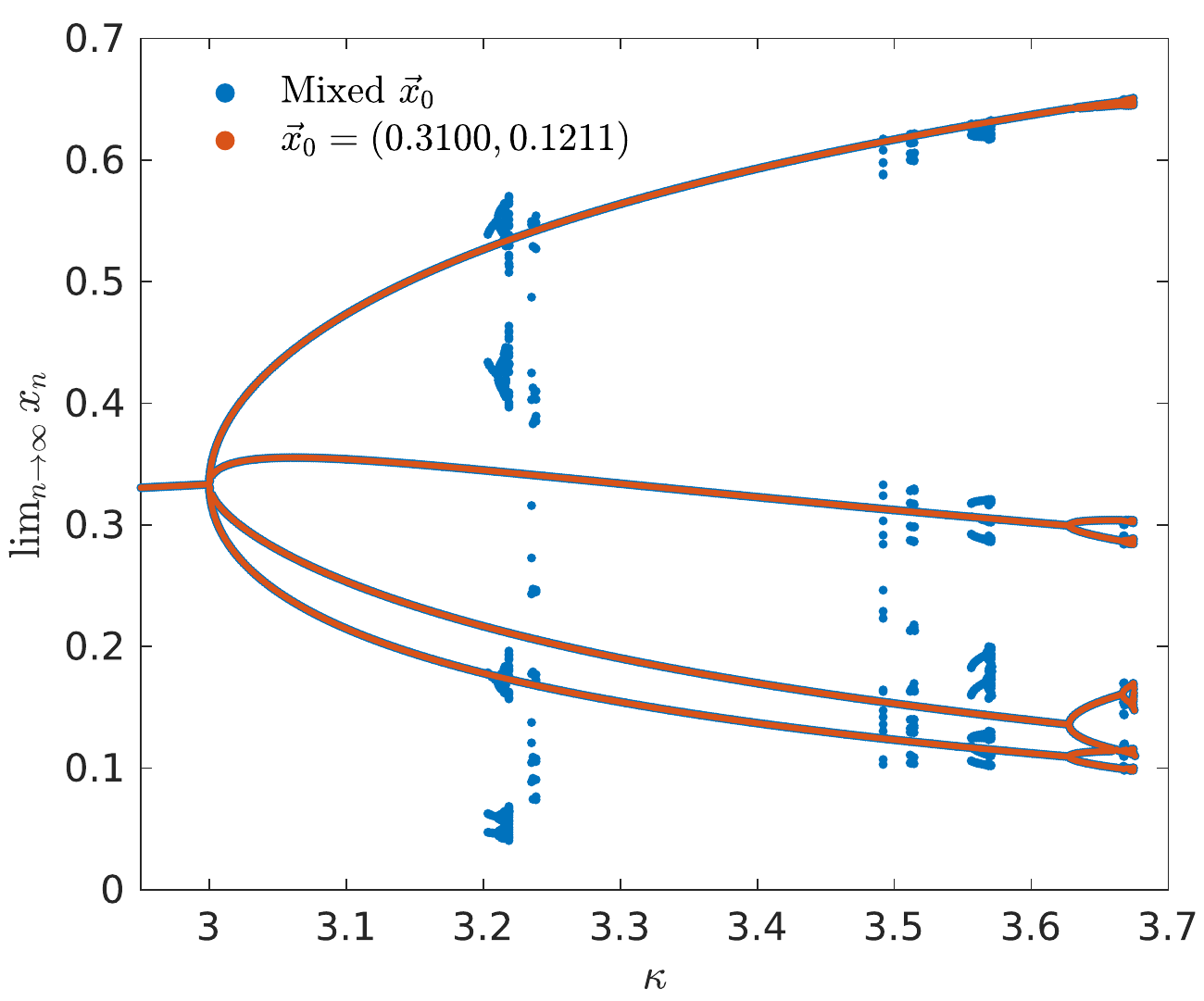}
	\end{subfigure}
	\begin{subfigure}{0.505\textwidth}
		\centering
		\includegraphics[width=\linewidth]{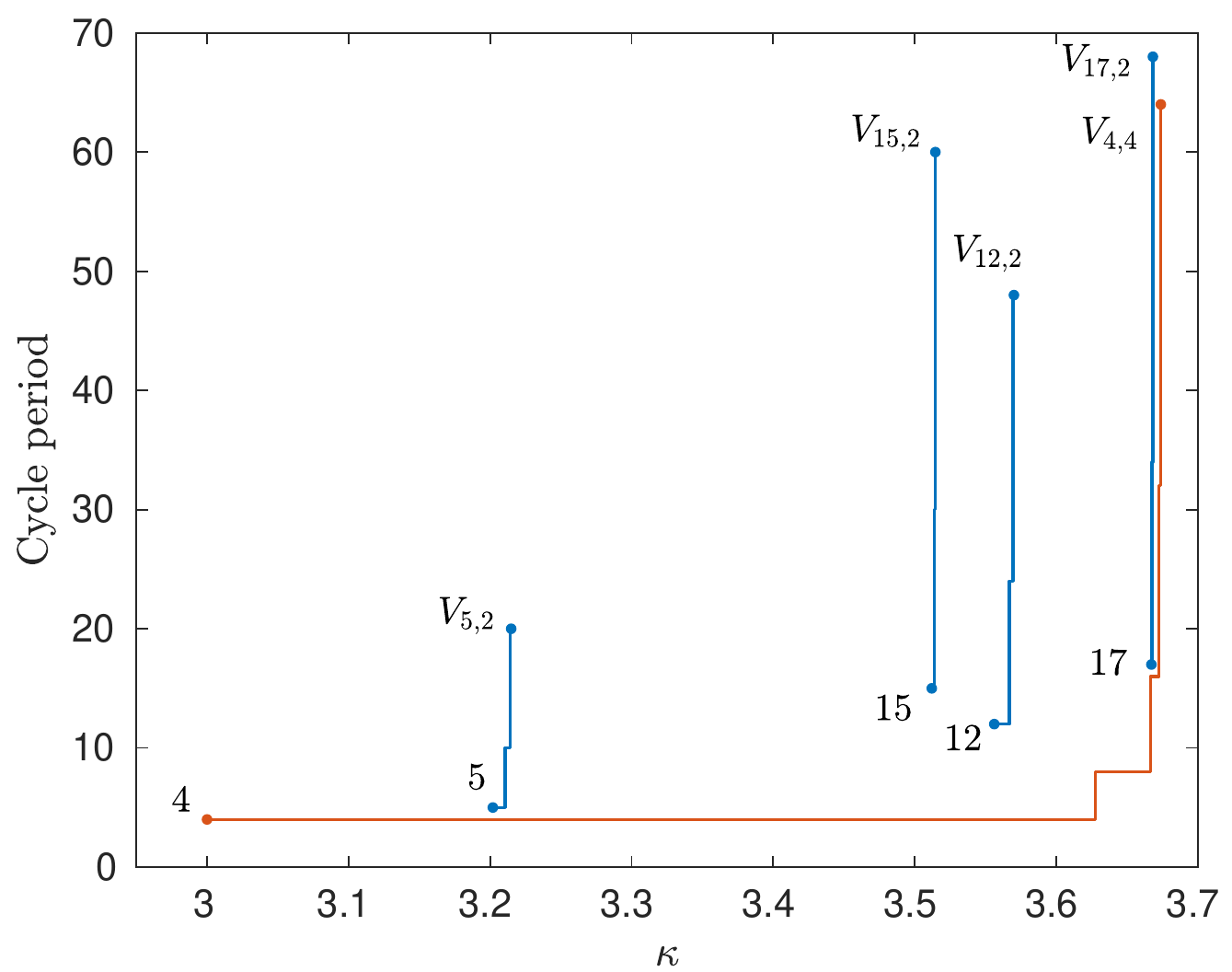}
	\end{subfigure}
	\caption{The bifurcation diagram of the $p_{\sf s}=0$ CBM (left) showing numerous saddle-node bifurcations up to $\kappa=\kappa_\mathrm{max}$ for a sampling of admissible initial conditions at each $\kappa$. Period-2 orbits are notably absent as the non-zero fixed point $x_+^*$ loses stability at $\kappa=3.0$ due to model constraints. No inadmissible $\kappa$ values are observed for the initial condition $\vec{x}_0=(0.3100,0.1211)$ up to $\kappa=\kappa_\mathrm{max}$. The corresponding branching diagram (right) shows the different periodic branches observed. The branch label $V_{j,k}$ corresponds to the branch starting from a period-$j$ orbit and which is observed to bifurcate $k$ times.}
	\label{BifurcationZoom}
\end{figure}

\begin{figure}
	\centering
	\includegraphics[width=\linewidth]{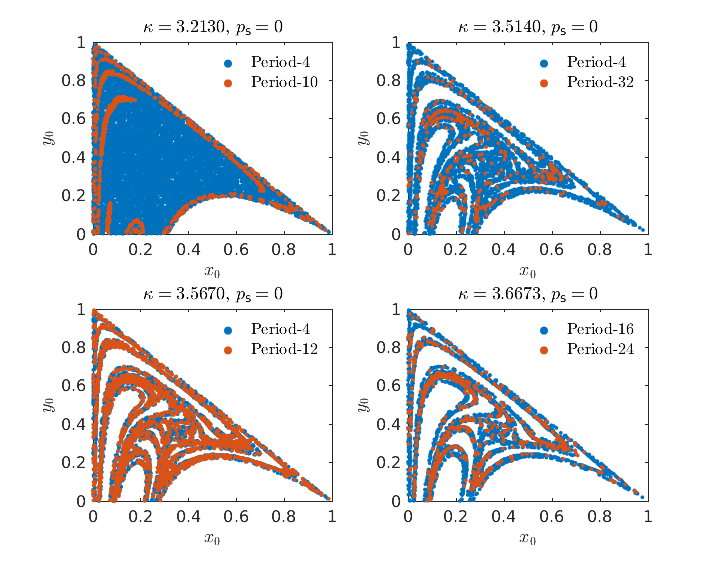}
	\caption{Basins of attraction corresponding to the select branches shown in Fig. \ref{BifurcationZoom} lead to windows of chaotic activity depending on the initial conditions and at observed values as low as $\kappa=\kappa_\mathrm{chaos}^{(5)}\approx3.2154$.}
	\label{basins}
\end{figure}

\noindent One route to chaos  begins with a period-doubling bifurcation, where the loss of stable fixed points is followed by period doubling starting with a period-2 orbit. To determine the stable period-2 orbits in the CBM, we need solutions to the squared map
\begin{align}
	\nonumber
	x_{n+2} &= F^{(2)}(x_n,y_n) = F(F(x_n,y_n))\\
	y_{n+2} &= G^{(2)}(x_n,y_n) = x_{n+1} = F(x_n,y_n),
\end{align}
such that $x\equiv x_{n+2}=x_n$ and $y\equiv y_{n+2}=y_n$. In solving this system, we must again ensure that $0\leq x_n+y_n\leq1$ for all $n$ and that the corresponding trajectories remain within the unit square. We find that when $p_{\sf s}=0$,
\begin{equation}
	y = \frac{1-x}{1+\kappa x}\kappa x,
\end{equation}
where there is a trivial solution, $x=0$, a real solution, 
\begin{equation}
	x =\frac{1}{3}+\left(-\frac{q}{2}+\sqrt{\frac{q^2}{4}+\frac{p^3}{27}}\right)^{1/3}+\left(-\frac{q}{2}-\sqrt{\frac{q^2}{4}+\frac{p^3}{27}}\right)^{1/3},
\end{equation}
where
$$
	p\equiv\frac{1-3\kappa(\kappa+1)}{3\kappa^3},
$$
and
$$
	q\equiv\frac{27(\kappa^2-1)-9\kappa^4(\kappa+1)-2\kappa^6}{27\kappa^6},
$$
and a complex-conjugate pair of solutions for $x$. This means that the only period-2 orbit possible within the constraints of the CBM is between the trivial solution and the real solution, however, because the trivial solution absorbs trajectories of the CBM when $p_{\sf s}=0$, there cannot be a period-2 orbit. We verify this numerically, finding a period-4 orbit following the loss of stability at $\kappa_\ast$ (cf. Fig. \ref{BifurcationZoom}). In addition to the primary bifurcation branch, shown in red, we observe additional branches, shown in blue, which arise depending on the choice of initial conditions. These branches amount to sudden transitions between stable orbits of different periods and lead to brief windows of chaos. Basins of attraction for stable orbits of different periods are shown in Fig. \ref{basins}, illustrating the variety in the asymptotic behavior depending on initial conditions. Similar behavior has previously been observed in the two-dimensional Hénon map \cite{mosekilde2000bifurcation}.

\subsection{Relation to the Hénon map}\label{henon}

\noindent It is interesting to remark that the CBM bears a striking resemblance to the well-known Hénon map \cite{Strogatz2015} with some crucial differences: the additional constraints on the CBM that the points $(x_n,y_n)$ remain in the unit square and the $x_ny_n$ cross-term which the Hénon map does not contain. The presence of the cross-term implies that the Jacobian of the map, about any point, is not a constant (it is equal to $|c x_n|$) and this property sets it apart from studies of the Hénon map and its generalizations in the literature such as \cite{hubbard1994henon, cai2015complex}. We can attempt to make contact with the vast literature on generalizations of the Hénon map by performing a change of variables. Indeed, since the map is quadratic, we may write $F_1(\vec{x}_n)$ from Eq. \ref{CBMtaur=2} in the following way:
\begin{equation}
	\label{F1}
	F_1(x_n,y_n) = p_{\sf s}+(c-p_{\sf s})x_n-p_{\sf s}y_n-\left(x_n\ y_n\right)\left(\begin{array}{cc} c & c/2\\ c/2 & 0\end{array}\right) \left(\begin{array}{c} x_n\\ y_n\end{array}\right).
\end{equation}
The $2\times 2$ matrix in this expression is symmetric, has constant coefficients, and can therefore be diagonalized by an orthogonal transformation to obtain an expression of the map where the non-linear terms are diagonal: 
\begin{equation}
	\label{diagonalCBM}
	\begin{array}{l}
		\displaystyle
		X_{n+1}\equiv F_1(X_n,Y_n)=p_{\sf s}+AX_n+BY_n+\lambda_1 X_n^2+\lambda_2Y_n^2\\
		Y_{n+1}\equiv F_2(X_n,Y_n)= X_n\cos\theta+ Y_n\sin\theta
	\end{array},
\end{equation}
where $\lambda_1=c \cos^2\theta-c\sin\theta\cos\theta$ and $\lambda_2=c \sin^2\theta+c\sin\theta\cos\theta$, and
\begin{equation}
	\label{CBMconstants}
	\begin{array}{l}
		\displaystyle
		A=c\cos\theta+p_{\sf s}\sin\theta \\
		B = c\sin\theta-p_{\sf s}\cos\theta
	\end{array}.
\end{equation}
However the resulting Jacobian, now dependent on the ``normal coordinates'' $X_n=x_n\cos\theta+y_n\sin\theta$ and $Y_n=y_n\cos\theta-x_n\sin\theta$, is $|p_{\sf s}+c^2 ((1+\sqrt{2})X_n\sin\theta+(1-\sqrt{2})Y_n\cos\theta)/2|$, where $\theta =(m-1/8)\pi$ or $\theta =(m+3/8)\pi$, with $m\in\mathbb{Z}$. The dependence on the angle $\theta$ suggests a rotation, which together with the constraints imposed by the CBM result in vortex-like structures which can be perceived in Fig. \ref{basins}. Note that $(X_n,Y_n)$ will still take bounded values, since $X_n^2+Y_n^2 = x_n^2+y_n^2$, however, the removal of the cross-term via orthogonal transformation still does not allow a direct relationship between the dynamics of the CBM map and those of the Hénon map. This seems to hint at the possibility that the CBM map belongs to a different universality class from the Hénon map.

\section{Conclusions}

\noindent We have mapped out  the route to chaos of the two-dimensional CBM with $k_{\sf in}=1$ and have provided evidence that the so-called \textit{quasiperiodic} phase of this model is, in fact, a chaotic phase. Our work highlights the level of care which must be taken when iterating the CBM maps.  The problem of determining admissible initial conditions and parameter values of the CBM represents a non-holomorphic generalization of the Julia and Mandelbrot sets, respectively. While promoting chaotic dynamics, the constraints of the CBM (expressed as relations between dynamical variables and boundary conditions) also eliminate period-2 trajectories while allowing others, a process amounting to a selection of frequencies which would be interesting to understand in terms of brain function. Moreover, the appearance of chaotic windows depending on initial conditions may represent regions of healthy brain activity, since they describe robustness under perturbations. Finally, despite their similarity, the Hénon map cannot be recovered from the CBM via orthogonal transformations. The presence of a non-constant Jacobian and restriction to the unit square are features of a novel class of generalized Hénon map. Our results thus suggest that the $k_{\sf in}=2$ CBM described in \cite{williams2014quasicritical} also belong to this class of generalized Hénon maps and feature unique and universal characteristics on its route to chaos. Finally, this study also highlights the wealth of the CBM class of models and indicates the interest for pursuing   investigation of the other models in the class. 

\section{Acknowledgments}
\noindent The authors would like to acknowledge support from the CNRS (Contract No. 1029456) which has made possible the residence of RVWG in the theoretical physics team of the Institut Denis Poisson (UMR7013).

\bibliographystyle{utphys}
\bibliography{CBM_v2}
\end{document}